\documentclass[twocolumn,amsmath,amssymb,prb,superscriptaddress]{revtex4-1}

\usepackage{graphicx}
\usepackage{dcolumn}
\usepackage{bm}
\usepackage{hyperref}
\usepackage{xcolor}
\usepackage{braket}

\allowdisplaybreaks
\usepackage{multirow}

\begin{document}

\title{Out of equilibrium density dynamics of a quenched fermionic system}
\author{S. Porta}
\affiliation{Dipartimento di Fisica, Universit\`{a} di Genova, Via Dodecaneso 33, I-16146 Genova, Italy}
\author{F. M. Gambetta}
\affiliation{Dipartimento di Fisica, Universit\`{a} di Genova, Via Dodecaneso 33, I-16146 Genova, Italy}
\affiliation{SPIN-CNR, Via Dodecaneso 33, I-16146 Genova, Italy}
\author{F. Cavaliere}
\affiliation{Dipartimento di Fisica, Universit\`{a} di Genova, Via Dodecaneso 33, I-16146 Genova, Italy}
\affiliation{SPIN-CNR, Via Dodecaneso 33, I-16146 Genova, Italy}
\author{N. Traverso Ziani}
\affiliation{Institute for Theoretical Physics and Astrophysics, University of W\"urzburg, D-97074 W\"urzburg, Germany }
\author{M. Sassetti}
\affiliation{Dipartimento di Fisica, Universit\`{a} di Genova, Via Dodecaneso 33, I-16146 Genova, Italy}
\affiliation{SPIN-CNR, Via Dodecaneso 33, I-16146 Genova, Italy}

\date{\today}


\begin{abstract}
Using a Luttinger liquid theory we investigate the time evolution of the particle density of a one-dimensional fermionic system with open boundaries and subject to a finite duration quench of the  inter-particle interaction. We provide analytical and asymptotic solutions to the unitary time evolution of the system, showing that {\em both} switching on and switching off the quench ramp create light-cone perturbations in the density. 
The post-quench dynamics is strongly affected by the interference between these two perturbations. In particular, we find that the discrepancy between the time-dependent density and the one obtained by a generalized Gibbs ensemble picture vanishes with an oscillatory behavior as a function of the quench duration, with local minima corresponding to a perfect overlap of the two light-cone perturbations. For adiabatic quenches, we also obtain a similar behavior in the approach of the generalized Gibbs ensemble density towards the one associated with the ground state of the final Hamiltonian.

\noindent PACS number(s): 71.10.Pm; 73.21.La; 67.85.Lm; 05.70.Ln
\end{abstract}

\maketitle

\section{Introduction}
The Hamiltonian $H(g)$, where $g$ is a set of parameters, of a generic quantum system can acquire a time dependence through the time dependence of the parameters $g$. In the past, time evolution for infinitely slow parameters variation has been extensively studied within the so called adiabatic approximation\cite{adiabatic,kato}, which is intimately related to the very rich Berry phase physics\cite{berry1,berry2,berrybook}. More recently, thanks to the advent of cold atoms\cite{cold1,cold2,cold3}, controllable time evolutions of the parameters, dubbed {\em quantum quenches}\cite{qq1,qq2}, could be experimentally addressed\cite{exp1,exp2,exp3,exp4,exp5,exp6,exp7,exp8}, boosting at the same time the theoretical inspection\cite{qq1,qq2,teo3,teo4,teo5,teo6,teo7}.\\
A remarkable result is represented by the sharp distinction between the behavior of non-integrable and integrable systems\cite{qq2}. While the first locally thermalize, despite the unitarity of the time evolution of the system after the quench, integrable models can reach, under proper conditions, states which are locally described by stationary non-thermal density matrices\cite{sred,nonint1,nonint2}. This peculiar behavior is associated with the presence of an infinite number of local or quasi local conserved quantities which prevent local thermalization.\\
In order to describe such a behavior, the concept of generalized Gibbs ensemble (GGE) has been introduced\cite{gge}.
Experimental evidence of such a thermodynamic ensemble has been gained by means of one dimensional (1D) bosonic atoms trapped in optical lattices realizing the so called Lieb-Lininger model\cite{expgge}.\\
From the theoretical point of view, the study of quantum quenches in integrable systems and their relaxation properties can benefit from the tools developed in order to deal with 1D quantum systems, among which the Luttinger liquid (LL) model\cite{voit,vonDelft:1998,giamarchi}.
Within the latter paradigm, the main effects of interactions can be included, while all correlation functions can be easily computed. The validity of LLs as low energy theories for 1D systems of fermions, bosons and spin, has been demonstrated experimentally by means of anomalous tunneling effects\cite{lu1,lu2}, or by observing spin-charge separation\cite{spincharge1,spincharge2,spincharge3}. Moreover, the LL theory represents a very useful tool for the study of a wide range of 1D systems, including integer\cite{iqhe} and fractional\cite{fqhe} quantum Hall effects and two dimensional topological insulators\cite{2dti1,2dti2,2dti3,2dti4}, weakly interacting quantum wires\cite{lu1}, even in the presence of spin-orbit coupling\cite{soc1,soc2,soc3,soc4,soc5,soc6}, carbon nanotubes\cite{lu2,luttingercnt1,luttingercnt2}, eventually including electron phonon coupling\cite{noi,grifoni}, spin chains\cite{luttingerspin1,luttingerspin2} and, complemented with its spin incoherent version\cite{fiete,mfg}, Wigner crystals\cite{wigmol1,wigmol2,wigmol3,wigmol4,wigmol5,wigmol6,wigmol7}.\\
The validity of the LL picture as a low energy theory for 1D Hamiltonians is however limited to the low energy excitations of gapless phases\cite{voit,giamarchi}. In quantum quench experiments its applicability is hence restricted to quenching protocols in which the Hamiltonian is gapless at every time and excitations with high energy are not significantly populated. The interesting cases of quantum quenches across a quantum phase transition\cite{qpt1,qpt2} are hence excluded.\\
Despite its limitations, quantum quenches in LLs define a very rich scenario\cite{qqreview} which is interesting by itself, and, through its simplicity, offers a perfect playground for understanding many general features. Sudden interaction quenches in LLs provided the first analytical confirmations to the GGE conjecture\cite{cazalilla} and allowed for very neat signatures of the (experimentally observed\cite{lightcone1}) light-cone effects\cite{lightcone2,lightcone3}, which are related to the Lieb-Robinson bounds\cite{liebrobinson1,liebrobinson2} for propagation of signals. Moreover the possibility of addressing different quenching protocols allowed a better understanding of the crossover from adiabatic to sudden quenches\cite{dora}. An even more interesting point about quantum quenches in LLs is that, when the LL picture is applicable, its predictions are, under some conditions of locality of the mapping to the LL\cite{universality2}, universal\cite{qqreview,universality1,universality2}: the long wavelength and low temperature results are in good agreement with numerical or semi-numerical results obtained by inspecting spin or fermionic lattice models. The power of the LL as a tool for studying 1D equilibrium systems is hence to some extent transferred to non-equilibrium situation.\\

\noindent In this paper we investigate the time evolution of the particle density of a fermionic spinless LL with open boundaries ~\cite{Fabrizio:1995} and subject to a finite duration linear quench of the inter-particle interaction. We focus on how the interplay between the time scale of the quenching protocol and the time scale introduced by the size of the system affects the transient and post-quench dynamics. 
Due to its finite size, the system will never reach a steady state but it will continue to oscillate around a certain configuration~\cite{Bocchieri:1957}. On the other hand, the finiteness in time of the quenching protocol has even more interesting consequences. Indeed, we show that {\em both} turning on and turning off of the quench ramp create light-cone perturbations in the density, which survive in the post-quench region. The two perturbations, starting from the edges~\cite{Manmana:2009}, travel ballistically through the system with the {\em instantaneous} velocity of the bosonic excitations of the LL~\cite{Bernier:2014} and bounce elastically whenever they reach a boundary. Furthermore, in the post-quench region, their wavefronts interfere destructively and, thus, the dynamics is strongly affected by their relative position inside the system. We find that for adiabatic quenches the post-quench average density oscillates in time around the one obtained with a GGE approach. For a fixed quench amplitude, the discrepancy between these two quantities decreases with an oscillatory behavior as a function of the quench duration, with local minima corresponding to a perfect overlap of the two light cones.  Finally we obtain that, for a long enough quench, the GGE density approaches the ground state density associated with the final Hamiltonian with the same decreasing oscillatory behavior as a function of the quench duration. \\

\noindent The paper is organized as follows. In Sec.~\ref{sec:model}  we introduce the open boundaries spinless LL model and in Sec.~\ref{subsec:bosonization} we solve analytically the time evolution dynamics of the bosonic operators for a linear quench of the inter-particle interaction with finite duration. We also examine the asymptotic solutions for sudden and adiabatic quenching protocols. Then, in Sec.~\ref{subsec:bosonization} we introduce the fermionic density operator, whose time-dependent average is evaluated in Sec.~\ref{subsec:averagerho}. In Sec.~\ref{results} our results are reported and discussed in details. Finally, Sec.~\ref{sec:conclusion} contains our conclusions. 

\section{The Model} \label{sec:model}
We consider a fermionic spinless LL confined in the region $ 0\leq x \leq L $ with open boundary conditions, subject to a quench of the inter-particle coulomb interaction. For $ t<0 $ the system is described by the initial Hamiltonian~\cite{voit,vonDelft:1998,giamarchi,Fabrizio:1995,density}
\begin{equation}
\label{eq:Hi}
H_i=\sum_{q>0}q\left[\left(v_F+\frac{g_{4,i}}{2\pi}\right)b^{\dagger}_qb_q-\frac{g_{2,i}}{4\pi}\left(b_qb_q+b^\dagger_qb^\dagger_q\right)\right]+\mathcal{E}(N),
\end{equation}
with momenta $ q=\pi n/L $ (with $ n $ a positive integer) and zero-mode energy $ \varepsilon(N)=\frac{\pi v_N}{2L}N^2 $, where $ v_N=v_F+(g_{2,i}+g_{4,i})/2\pi $ and $ N $ is the fermionic number operator. In the following we will focus on the case $ g_{2,i}=g_{4,i}=g_i $. The Hamiltonian of Eq.~\eqref{eq:Hi} can be brought in diagonal form by means of a Bogoliubov transformation, 
\begin{equation}
\label{eq:Hi_diag}
H_i=\sum_{q>0}qv_i\beta^{\dagger}_q\beta_q+\mathcal{E}(N). 
\end{equation}
Here, $ \beta_q=u_+b_q+u_-b_q^{\dagger} $, with $ u_{\pm}=(K_i^{1/2}\pm K_i^{-1/2})/2 $, $ v_i=v_F/K_i $ is the velocity of the bosonic modes and $ K_i=(1+g_i/\pi v_F)^{-1/2} $ is the initial Luttinger parameter, describing the intensity of inter-particle interaction. In particular, $ 0<K_i<1 $ for repulsive interaction and $ K_i=1 $ for non-interacting particles. 
At $ t=0 $ the system undergoes a quench of the inter-particle interaction~\cite{cazalilla,Iucci:2009,dora,Sachdeva:2014} from $ g_i $ to $ g_f $ with a general quenching protocol $ Q(t) $ with time duration $ \tau $ and amplitude $ \delta_g=g_f-g_i=v_F(K_f^{-2}-K_i^{-2})/2 $, such that $ Q(t\leq0)=0 $ and $ Q(t\geq\tau)=1 $. Here, $ K_f=(1+g_f/\pi v_F)^{-1/2} $ is the final Luttinger parameter. 
In terms of the $ \beta_q $ operators, the  time-dependent Hamiltonian is
\begin{align}
H(t)&=\sum_{q>0}q\left[v(t)\beta^{\dagger}_q\beta_q+\frac{1}{2}g(t)\left(\beta^{\dagger}_q\beta^{\dagger}_q+\beta_q\beta_q\right)\right]\nonumber\\
&+\mathcal{E}(N)+\Omega(t),\label{eq:Ht}
\end{align}
with
\begin{align}
\Omega(t) &=\frac{e^{\pi\alpha/L}}{2(1-e^{\pi\alpha/L})^2}Q(t)\delta_g(K_i-1)\\
v(t)&=v_i+K_i\delta_gQ(t),\label{eq:vt}\\
g(t)&=-K_i\delta_gQ(t)\label{eq:gt}.
\end{align}
Here, $ \Omega(t) $ is the energy mismatch of the instantaneous ground state with respect to the initial one and   $ \alpha $ is the shortest-length cutoff of the  theory. The time evolution of the bosonic operator can be obtained from the Heisenberg equation of motion~\cite{dora,Sachdeva:2014},
\begin{equation}
\label{eq:betaevolution}
\beta_q(t)=f(q,t)\beta_q+h^*(q,t)\beta_q^{\dagger},
\end{equation}
with the bosonic operator on the right hand side evaluated at $ t=0 $.  The coefficients $ f(q,t) $ and $ h(q,t) $ satisfy the relation $ |f(q,t)|^2-|h(q,t)|^2=1\ \forall q,t $ and the system of coupled differential equations~\cite{dora,Sachdeva:2014}
\begin{equation}
\label{eq:desystem}
i\begin{bmatrix}
\dot{f}(q,t)\\
\dot{h}(q,t)
\end{bmatrix}=q\begin{bmatrix}
v(t) & g(t)\\
-g(t) & -v(t)
\end{bmatrix}
\begin{bmatrix}
f(q,t)\\
h(q,t)
\end{bmatrix}
\end{equation}
with initial condition $ f(q,0)=1 $ and $ h(q,0)=0 $. Here, the dot indicates the derivate with respect to time.  To solve this system it is convenient to shift to the  basis defined by the functions
\begin{subequations}
\label{eq:DandS}
\begin{align}
D(q,t)&=f(q,t)-h(q,t),\\
S(q,t)&=f(q,t)+h(q,t).
\end{align}
\end{subequations}
In this new basis, using the explicit expressions for $ v(t) $ and $ g(t) $ given in Eqs.~(\ref{eq:vt}, \ref{eq:gt}), one can rewrite the original system of Eq. \eqref{eq:desystem} as
\begin{subequations}
\label{eq:desystems2}
\begin{align}
\ddot{D}(q,t)+\left[q\bar{v}(t)\right]^2D(q,t)&=0,\label{eq:deD}\\
\dot{D}(q,t)+iqv_iS(q,t)&=0,
\end{align}
\end{subequations}
with initial conditions $ D(q,0)=1 $ and $ \dot{D}(q,0)=-iqv_i $. Here, 
\begin{equation}
\label{eq:vinstgen}
\bar{v}(t)=v_i\sqrt{1+\frac{2K_i\delta_g}{v_i}Q(t)}
\end{equation}
is the {\em instantaneous } bosonic modes velocity. Indeed, at any time $ \bar{t} $ the system Hamiltonian of Eq.~\eqref{eq:Ht} is diagonal in the instantaneous bosonic operators~\cite{Bernier:2014} $ \beta_{q,\bar{t}} $, 
\begin{equation}
H(\bar{t})=\sum_{q>0}q\bar{v}(\bar{t})\beta_{q,\bar{t}}^{\dagger}\beta_{q,\bar{t}},
\end{equation}
with $ \bar{v}(\bar{t}) $ given precisely by Eq.~\eqref{eq:vinstgen}. Here, $ \beta_{q,\bar{t}}=u_{\bar{t},+}b_q+u_{\bar{t},-}b_q^{\dagger} $, with $ u_{\bar{t},\pm}=[(\mu(\bar{t}))^{-1/2}\pm (\mu(\bar{t}))^{1/2}]/2 $ and $ \mu(\bar{t})=[1+2K_i\delta_gQ(\bar{t})/v_i]^{1/2} $.

\noindent Once solved Eq.~\eqref{eq:deD} for the function $ D(q,t) $, one immediately obtains $ S(q,t) $ by differentiation and then, from Eq.~\eqref{eq:DandS}, the coefficients $ f(q,t) $ and $ h(q,t) $. Thus, in the following we will focus only on the resolution of Eq.~\eqref{eq:deD}. 
\subsection{Linear quench}
\noindent The model developed above is valid for any arbitrary quenching protocol $ Q(t) $ with $ Q(t\leq0)=0 $ and $ Q(t\geq\tau)=1 $. We now specialize the discussion by choosing a quench with a linear ramp~\cite{dora,Sachdeva:2014}:
\begin{equation}
\label{eq:protocol}
Q(t)=\begin{cases}
0 &\text{for }t<0\text{ (region I)},\\
t/\tau &\text{for }0\leq t\leq\tau\text{ (region II)},\\
1 &\text{for }t>\tau\text{ (region III)}.
\end{cases}
\end{equation}
With this protocol Eq.~\eqref{eq:deD} for $ D(q,t) $ can be solved in the three regions introduced in Eq.~\eqref{eq:protocol}, with properly matching conditions on the boundaries of each region. We obtain
\begin{widetext}
\begin{subequations}
\label{eq:D}
\begin{align}
D^{\mathrm{I}}(q,t)&=e^{-i v_i t},\\
D^{\mathrm{II}}(q,t)&=\frac{\pi \Delta_q\sqrt{1+\eta t/\tau}}{\sqrt{3}}\left\{ \mathcal{A}(\Delta_q)J_{-\frac{1}{3}}\left[\Delta_q(1+\eta t/\tau)^{3/2}\right]+\mathcal{B}(\Delta_q)J_{\frac{1}{3}}\left[\Delta_q(1+\eta t/\tau)^{3/2}\right]\right\},\label{eq:DII}\\
D^{\mathrm{III}}(q,t)&=\frac{\pi \Delta_q\mu}{\sqrt{3}} \left\{\mathcal{C}(\Delta_q,\mu)\cos\left[qv_f\left(t-\tau\right)\right]-\mathcal{S}(\Delta_q,\mu)\sin\left[qv_f\left(t-\tau\right)\right]\right\} . \label{eq:DIII}
\end{align}
\end{subequations}
\end{widetext}
Here, $ J_\nu(x) $ are Bessel functions of the first kind of order $ \nu $. Furthermore, 
\begin{align}
\mathcal{A}(\Delta_q)&=J_{-\frac{2}{3}}(\Delta_q)+ iJ_{\frac{1}{3}}(\Delta_q),\\
\mathcal{B}(\Delta_q)&=J_{\frac{2}{3}}(\Delta_q)- iJ_{-\frac{1}{3}}(\Delta_q),\\
\mathcal{C}(\Delta_q,\mu)\!&=\!\mathcal{A}(\Delta_q)J_{-\frac{1}{3}}\left(\Delta_q\mu^{3}\right)\!+\!\mathcal{B}(\Delta_q)J_{\frac{1}{3}}\left(\Delta_q\mu^{3}\right),\\
\mathcal{S}(\Delta_q,\mu)\!&=\!\mathcal{A}(\Delta_q)J_{\frac{2}{3}}\left(\Delta_q\mu^{3}\right)\!-\!\mathcal{B}(\Delta_q)J_{-\frac{2}{3}}\left(\Delta_q\mu^{3}\right),
\end{align}
and we have introduced the parameters
\begin{equation}
\label{eq:parameters}
\Delta_q=\frac{2}{3}\frac{qv_i\tau}{\eta},\quad\eta=\mu^2-1,\quad \mu=\frac{K_i}{K_f}.
\end{equation}
Although Eq.~\eqref{eq:D} provides the exact solution of the non-equilibrium problem over the whole range of time, in order to get more physical insight it is convenient to analyze the two opposite limits of sudden and adiabatic quench~\cite{cazalilla,qqreview,dora}. These regimes are controlled by the value of the parameter $ \Delta_q$ introduced in Eq.~\eqref{eq:parameters}. In particular, we have that $ \Delta_q\ll1 \ \forall q $ defines the sudden quench limit, while the condition $ \Delta_q\gg1  \ \forall q$ determines the adiabatic one. Since $ \Delta_q\propto q $, the validity of the sudden approximation strongly depends on the highest-momentum cutoff of the theory $ q_c $, while the adiabatic one is cutoff independent. For a quench of fixed amplitude $ \delta_g $, the above conditions can be recast as functions of the ramp time duration
\begin{align}
\tau\ll\tau_{\mathrm{sq}}\equiv\frac{3|\eta|}{2 q_c v_i} \quad&\rightarrow\quad \text{sudden quench limit},\\
\tau\gg\tau_{\mathrm{ad}}\equiv\frac{3L|\eta|}{2\pi v_i} \quad&\rightarrow\quad \text{adiabatic quench limit}.\label{eq:adcondition} 
\end{align}
A physical interpretation of the two conditions will be given later (see Sec.~\ref{results}). 
The value of $ \tau_{\mathrm{sq}} $ and $ \tau_{\mathrm{ad}} $ in real systems depends on their nature; for example~\cite{footnote:tauvalues}, for a carbon nanotube one has $ \tau_{\mathrm{ad}}\sim10^{-13}|\eta|\ \mathrm{s} $ and $ \tau_{\mathrm{sq}}\sim10^{-14}|\eta|\ \mathrm{s} $, while in a cold fermionic gas $ \tau_{\mathrm{ad}}\sim10^{-7}|\eta|\ \mathrm{s} $ and $ \tau_{\mathrm{sq}}\sim10^{-8}|\eta|\ \mathrm{s} $. Since for a not too strong quench $ |\eta|\sim1 $, from the above values the importance of studying adiabatic quenches should be evident, especially in the solid state realm.

\noindent {\em Sudden quench limit.} In the sudden quench case~\cite{cazalilla,Iucci:2009,qqreview}, expanding Eq.~\eqref{eq:DIII} for $ \tau\rightarrow0 $, i.e. to the zeroth order in $ \Delta_q $, one obtains (for $ t>0^{+} $ )
\begin{equation}
\label{eq:DIIISQ}
D^{\mathrm{III}}_{\mathrm{sq}}(q,t)=\cos[qv_f(t-\tau)]-i\mu^{-1}\sin[qv_f(t-\tau)].
\end{equation}

\noindent{\em Adiabatic quench limit.} On the other hand, in the limit $ \tau\gg\tau_{\mathrm{ad}} $ an asymptotic expansion~\cite{Murray:1984} to the first order in $ \Delta_q^{-1} $ of Eqs.~(\ref{eq:DII}, \ref{eq:DIII}) gives
\begin{align}
\label{eq:DIIad}
D^{\mathrm{II}}_{\mathrm{ad}}(q,t)&\approx\frac{1}{(1+\eta t/\tau)^{1/4}}\left\{ \cos\left[\Delta_q\left((1+\eta t/\tau)^{3/2}-1\right)\right]\right.\nonumber\\
&\left.+\left(\frac{1}{6\Delta_q}-i\right)\sin\left[\Delta_q\left((1+\eta t/\tau)^{3/2}-1\right)\right]\right\}
\end{align}
and
\begin{align}
D^{\mathrm{III}}_{\mathrm{ad}}(q,t)&\approx\frac{1}{24\mu^{5/2}qv_{f}}\left\{ \mathcal{C}_{\mathrm{ad}}(\Delta_q,\mu)\cos\left[qv_{f}(t-\tau)\right]\right.\nonumber\\
&-\mathcal{S}_{\mathrm{ad}}(\Delta_q,\mu)\sin\left[qv_{f}(t-\tau)\right],\label{eq:DIIIad}
\end{align}
respectively. Here,
\begin{align}
\mathcal{C}_{\mathrm{ad}}(\Delta_q,\mu)\!&=\!4\mu^{2}qv_{f}\left[6\cos\bar{\Delta}_q\!+\!(\Delta_q^{-1}\!-\!6i)\sin\bar{\Delta}_q\right],\\
\mathcal{S}_{\mathrm{ad}}(\Delta_q,\mu)\!&=\!\frac{\eta\Delta_q}{\tau}\left[6\left(\Delta_q^{-1}-\mu\Delta_q^{-1}+6i\mu^{3}\right)\cos\bar{\Delta}_q\right.\nonumber\\
&\left.+6\left(6\mu^{3}-i\Delta_q^{-1}\right)\sin\bar{\Delta}_q\right],
\end{align}
and $ \bar{\Delta}_q=\Delta_q(\mu^3-1) $.
\subsection{Bosonization and density operator}\label{subsec:bosonization}
\noindent Following the standard bosonization prescriptions~\cite{voit,vonDelft:1998,giamarchi}, we decompose the fermionic field $ \Psi(x) $ in right-- ($ R(x) $) and left--  ($ L(x) $) moving fields. Working in the Heisenberg picture, we write
\begin{equation}
\Psi(x,t)=R(x,t)+L(x,t).
\end{equation}
The open boundary conditions~\cite{Fabrizio:1995} imply that $ L(x,t)=-R(-x,t) $, with the bosonized right--moving field given by
\begin{equation}
\label{eq:bosR}
R(x,t)=\frac{F(t)}{\sqrt{2\pi\alpha}}e^{ik_Fx/L}e^{i\Phi(x,t)}.
\end{equation}
Here, $ k_F=\pi N/L $ is the Fermi momentum, 
\begin{align}
	\label{eq:phi}
	\Phi(x,t)&=\frac{1}{\sqrt{K_i}}\sum_{q>0}\sqrt{\frac{\pi}{L q}}e^{iqx-\alpha q/2}\nonumber\\
	&\times[\cos(qx)+iK_i\sin(qx)]\beta_q(t)+\mathrm{h.c.}
\end{align}
is the bosonic field, with $ \beta_q(t) $ given in Eq.~\eqref{eq:betaevolution}, and $ F(t) $ is the time-evolved Klein factor
\begin{equation}
F(t)=e^{-i\frac{\pi v_N}{2L}(2N+1)t}F,
\end{equation}
with $ F $ the Klein factor at $ t=0 $, obtained using the commutation relation $ [N,F]=-F $.

\noindent The particle density operator of the system is given by~\cite{Fabrizio:1995,density}
\begin{equation}
\label{eq:rho}
\rho(x,t)=\rho_{\mathrm{LW}}(x,t)+\rho_{\mathrm{F}}(x,t),
\end{equation}
with
\begin{subequations}
\begin{align}
\rho_{\mathrm{LW}}(x,t) &=\rho_{\mathrm{R}}(x,t)+\rho_{\mathrm{R}}(-x,t),\\
\rho_{\mathrm{F}}(x,t) &=-R^{\dagger}(-x,t)R(x,t)-R^{\dagger}(x,t)R(-x,t)
\end{align}
\end{subequations}
the long-wave and Friedel contributions, respectively. Here,
\begin{equation}
\rho_{\mathrm{R}}(x,t)=\frac{N}{2L}-\frac{\partial_x \Phi(x,t)}{2\pi}
\end{equation}
is the right--moving fermionic density. Using the bosonized field of Eq.~\eqref{eq:bosR}, the Friedel term can be written as
\begin{equation}
\label{eq:rhoFriedel}
\rho_{\mathrm{F}}(x,t)=-\frac{1}{\pi\alpha}\cos\left[2k_Fx-2\Phi_a(x,t)-2f(x)\right],
\end{equation}
where
\begin{align}
\label{eq:phi_a}
\Phi_a(x,t)&=\frac{1}{2}\left[\Phi(-x,t)-\Phi(x,t)\right],\\
f(x)&=\frac{1}{2}\arctan\left[\frac{\sin(2\pi x /L)}{e^{\alpha \pi/L}-\cos(2\pi x/L)}\right].
\end{align}
\subsection{Average fermionic density}\label{subsec:averagerho}
Working in the zero-temperature limit and assuming the system prepared in the ground state of $ H_i $ for $ t<0 $, we can now calculate the average fermionic density. The long-wave term simply evaluates $ N/L $, while the Friedel one can be obtained plugging the bosonic fields of Eqs.~(\ref{eq:phi}, \ref{eq:phi_a}) in Eq.~\eqref{eq:rhoFriedel} and following the standard bosonization procedure~\cite{voit,vonDelft:1998,giamarchi,Fabrizio:1995,density}. The final result is
\begin{equation}
\label{eq:avrho}
\langle\rho(x,t)\rangle_i=\frac{N}{L}\left\{ 1-E(x,t)\cos\left[2k_Fx-2f(x)\right] \right\},
\end{equation}
where $ \langle...\rangle_i  $ represents the quantum average over the ground state of $ H_i $. Interestingly, the time dependence is contained only in the {\em envelope function}
\begin{equation}
\label{eq:envelope}
E(x,t)\!=\!\exp\left[\!-\frac{2\pi K_i}{L}\sum_{q>0}\frac{e^{-\alpha q}}{q}\sin^2(qx)|D(q,t)|^2\right].
\end{equation}
For $t>\tau$ the envelope function, along with the density, is a periodic function with period $ \mathcal{T}=L/v_{f} $, i.e. $E(x,t+\mathcal{T})=E(x,t)$. This is a direct consequence of the so called quantum recurrence theorem for systems with a discrete energy spectrum~\cite{Bocchieri:1957}. 

\noindent We recall that for a standard non-quenched spinless LL with Luttinger parameter $ K_{j} $ one has $ |D(q,t)|^2=1 \ \forall q,t$. Thus, the non-quenched average fermionic density $ \langle\rho_{\mathrm{nq}}(x)\rangle $ is time-independent, with an envelope function given by~\cite{density}
\begin{equation}
\label{eq:Estd}
E_{\mathrm{nq},j}(x)=\left[\mathcal{K}(x)\right]^{K_j},
\end{equation}
where
\begin{align}
\label{eq:K}
\mathcal{K}(x)&=\exp\left[-\frac{2\pi}{L}\sum_{q>0}\frac{e^{-\alpha q}}{q}\sin^2(qx)\right]\nonumber\\
&=\frac{\sinh(\pi\alpha/2L)}{\sqrt{\sinh^2(\pi\alpha/L)+\sin^2(\pi x/L)}}.
\end{align}

\section{Results}\label{results}
In this section we analyze the behavior of the average density obtained in Eq.~\eqref{eq:avrho} for the quenched system in the different regimes. For definiteness, we focus on quenching protocols with $K_{f}<K_{i}$: we have checked that the opposite situation leads to qualitatively similar behavior. Since the average density has the general form given in Eq.~(\ref{eq:avrho}), it is clear that all the dynamics is captured by the envelope function $E(x,t)$, which will then constitute the focus of this section.

In order to better highlight the dynamics of the envelope function $E(x,t)$, we will begin by studying $\delta E(x,t)=E(x,t)-E_{\mathrm{nq},i}(x)$ where $E_{\mathrm{nq},i}(x)$ represents the standard steady envelope function of the density for a non-quenched LL in the initial state with interaction parameter $K_i$ -- see Eq.~\eqref{eq:Estd}.
\begin{figure}[htbp]
	\begin{center}
		\includegraphics[width=0.95\columnwidth]{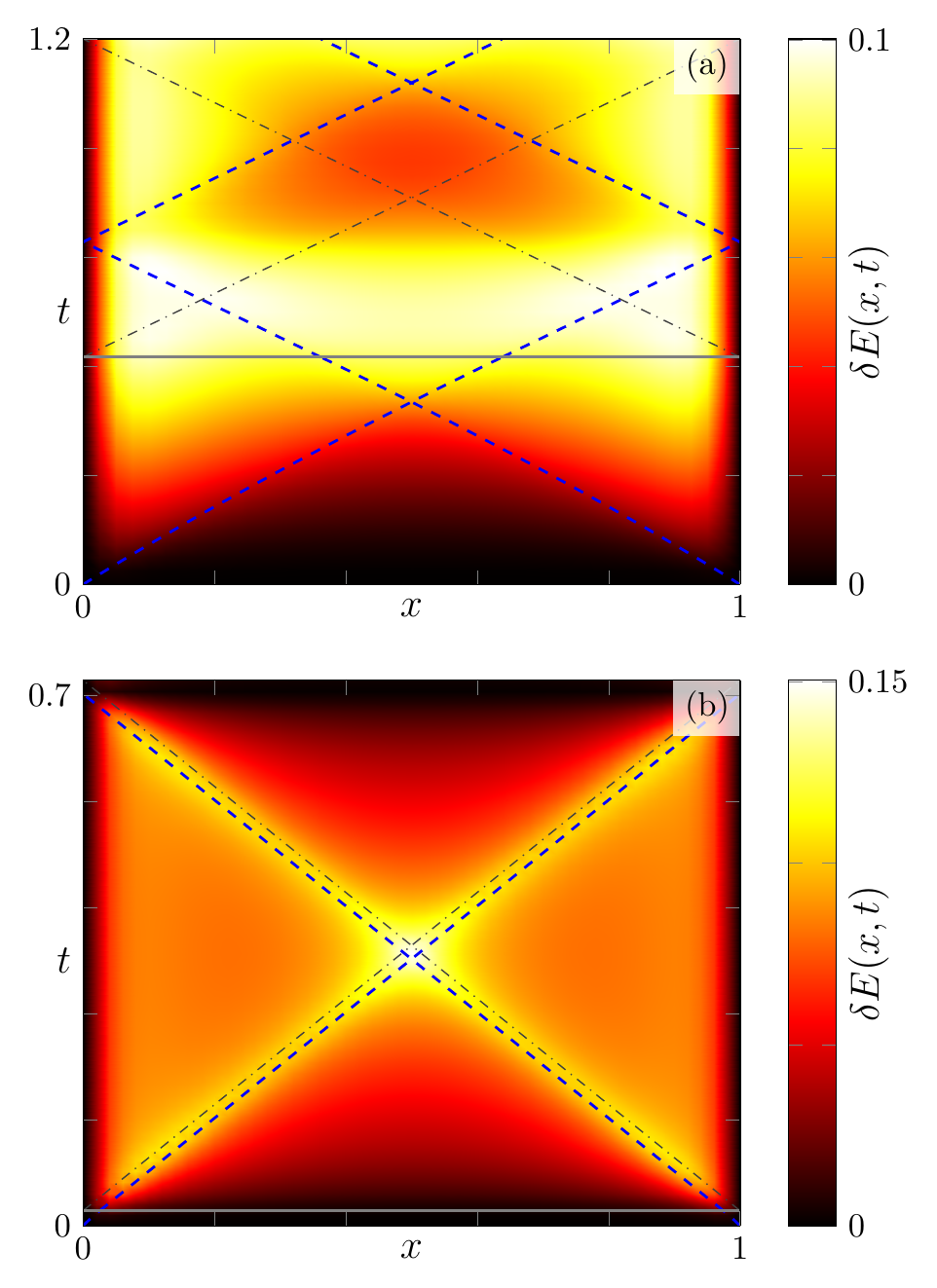}
		\caption{(Color online) Density plot of $ \delta E(x,t)=E(x,t)-E_{\mathrm{nq},i}(x) $ as a function of $ x $ (units $ L $) and $ t $ (units $ L/v_F $) for a quench from $ K_i=0.9 $ to $ K_f=0.7 $. The dashed blue lines represent LC1, while the dash-dotted black ones highlight LC2. The solid gray line denotes the end of the ramp. Ramp time: $ \tau=0.5 $ (Panel (a)) and $ \tau=0.02 $ (Panel (b)) (units $L/v_F  $). Here, $ \alpha/L=0.05 $. }
		\label{fig:Fig1}
	\end{center}
\end{figure}
Figure~\ref{fig:Fig1} shows $\delta E(x,t)$ for a quench from $ K_i=0.9 $ to $ K_f=0.7 $ for two opposite situations. In Panel (a) $\tau>\tau_{\mathrm{ad}}$  has been chosen, corresponding to an adiabatic quench, while in Panel (b) $\tau\ll\tau_{\mathrm{sq}}$ in the sudden quench limit has been employed. In the adiabatic case two phases can be clearly distinguished: as soon as the quench begins, in addition to an overall growth of the envelope function over time, a perturbation starts propagating from the edges through the system. Since the edges play here a special role breaking the translational invariance of the system, it is then natural that quench excitations begin their journey from these points. This identifies a first ``light cone" (LC1) which travels in an accelerated fashion for $0\leq t\leq \tau$ -- see the dashed lines in Fig.~\ref{fig:Fig1} highlighting the wavefront of the perturbation. In the post-quench regime, for $t>\tau$, a second LC (LC2) emerges from the edges -- see the dash-dotted lines in Fig.~\ref{fig:Fig1}  -- and both propagate at constant velocity $v_{f}$. The two light cones travel ballistically through the system and bounce elastically whenever they reach one of its boundary. The same situation formally occurs also in the case of a sudden quench, depicted in Panel (b), only that to all extents LC1 and LC2 merge and the ramp dynamics is indistinguishable due to its shortness. We will now proceed discussing the two limits in details.

\subsection{Sudden quench}
In the sudden quench limit $\tau\to 0$, from Eqs.~(\ref{eq:DIIISQ},\ref{eq:envelope}), the following analytic expression for $E(x,t)$ is found
\begin{equation}
\label{eq:ESQ}
E(x,t)=E_{\mathrm{sq}}^{\mathrm{(GGE)}}(x)\left[\frac{\mathcal{K}(v_{f}t)}{\sqrt{\mathcal{K}(x-v_{f}t)\mathcal{K}(x+v_{f}t)}}\right]^{\nu}\, ,
\end{equation}
where $E_{\mathrm{sq}}^{\mathrm{(GGE)}}(x)$, defined in Eq.~(\ref{app:eq:ESQGGE}), is the envelope function obtained assuming for the system a density matrix within the GGE and $\nu=(K_{f}^{2}-K_{i}^{2})/(2K_{i}^{2})$. 
\begin{figure}[htbp]
	\begin{center}
		\includegraphics[width=\columnwidth]{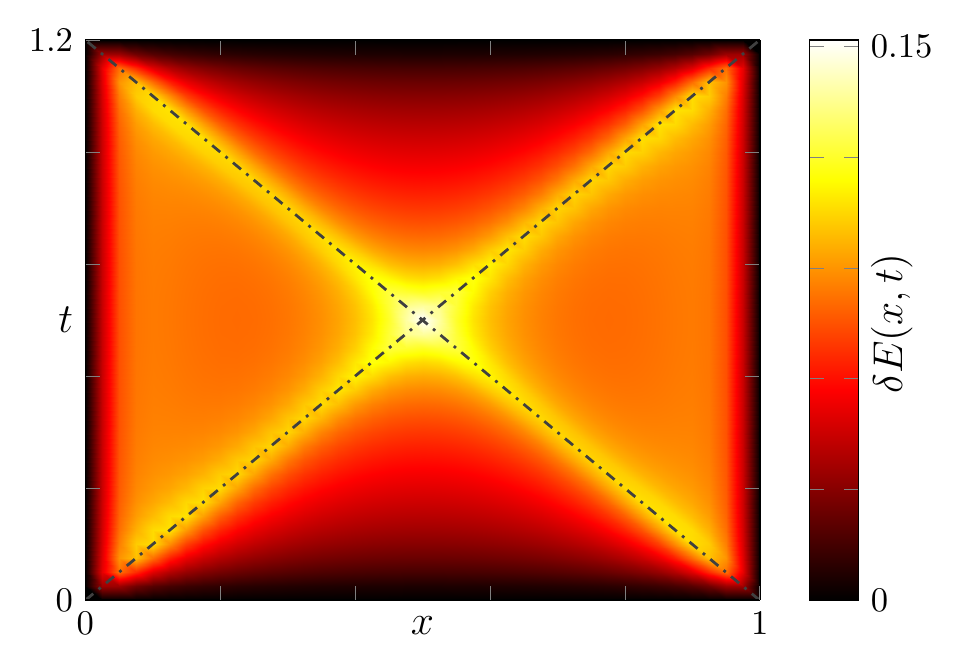}
		\caption{(Color online) Same as in Fig.~\ref{fig:Fig1} for a sudden quench from $ K_i=0.9 $ to $ K_f=0.7 $. Note that only one light cone is present. }
		\label{fig:Fig6}
	\end{center}
\end{figure}

\noindent From the denominator of the second factor of Eq.~\eqref{eq:ESQ}, it follows that in this regime LC1 and LC2 merge and only a single light-cone perturbation with counter-propagating branches moving at constant velocity $ v_f $ can be detected, as shown for instance in the example of Fig.~\ref{fig:Fig6}.
\begin{figure}[htbp]
	\begin{center}
		\includegraphics[width=\columnwidth]{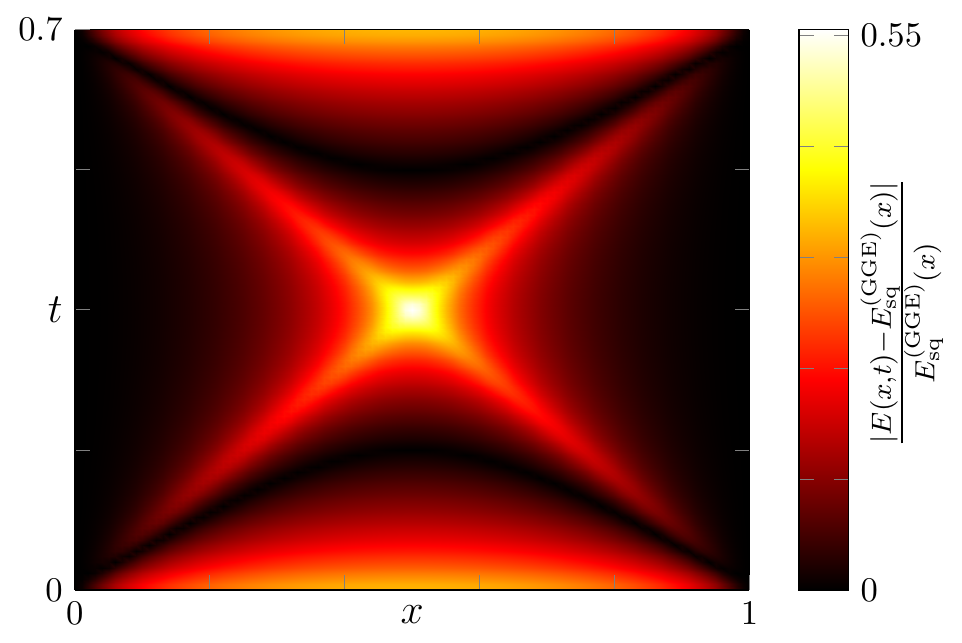}
		\caption{(Color online) Density plot of the relative difference $|E(x,t)-E_{\mathrm{sq}}^{\mathrm{(GGE)}}(x)|/E_{\mathrm{sq}}^{\mathrm{(GGE)}}(x)$ as a function of $ x $ (units $ L $) and $ t $ (units $ L/v_F $) for a sudden quench from $ K_i=0.9 $ to $ K_f=0.7 $. Here, $ \alpha/L=0.05 $. }
		\label{fig:Fig7}
	\end{center}
\end{figure}

\noindent In the sudden quench limit the envelope function $E(x,t)$ fluctuates quite largely around the GGE result except for the regions $x\ll L/2$ and $L-x\ll L/2$ around the edges where, as discussed in Appendix~\ref{subsec:GGE}, $E(x,t)\to E_{\mathrm{sq}}^{\mathrm{GGE}}(x)$. In the rest of the system, however, the GGE fails pretty badly in capturing the density dynamics -- see Fig.~\ref{fig:Fig7}. 
\subsection{Adiabatic quench}
More articulated is the regime of an adiabatic quench. Here we distinguish between two different regimes, namely (1) the ``on-ramp" phase with $0\leq t\leq\tau$ and (2) the post-quench phase for $t>\tau$.\\

\noindent (1) {\em{On-ramp}}. Besides an overall growth of the envelope function over time, this phase is governed by the dynamics of LC1, which can be readily understood when observing that during the ramp, the excitations at time $t$ are harmonic bosons with instantaneous velocity $\bar{v}(t)=v_{i}\sqrt{1+\eta t/\tau}$ -- see Eq.~\eqref{eq:vinstgen}. Therefore, as the system is non-dispersive, it is reasonable to assume that overall the perturbation propagates with the same instantaneous velocity. Since the perturbation detaches from the edges, we can readily find the distance $\ell(t)$ traveled by it considering the equation of motion 
\begin{equation}
\label{eq:EOM}
\dot{\ell}(t)=\pm\bar{v}(t)\, ,	
\end{equation}
where the sign $ + $ ($ - $) refers to the propagation to the left (to the right). The perturbation front originating at $ x=0$ propagates to the right until it eventually reaches $x=L$. In the meanwhile, the front originating at $x=L$ propagates to the left (with the same modulus velocity) until it reaches $x=0$. Then, the motion eventually repeats. Since the modulus of the velocities is the same, it follows that the total distance traveled by each of the wavefronts in the time $t$ can be obtained integrating Eq.~(\ref{eq:EOM}) with the sign + and boundary condition $\ell(0)=0$,
\begin{equation}
\label{eq:dist1}
\ell(t)=\ell_0\left[\left(1+\frac{\eta t}{\tau}\right)^{\frac{3}{2}}-1\right]\, ,
\end{equation}
with $\ell_0=2v_{i}\tau/3\eta$. Note that the adiabatic condition in Eq.~\eqref{eq:adcondition} can be recast as $\ell_0\gg L$. Of particular interest is the total distance $d=\ell(\tau)$ traveled by the LC1 fronts during the ramp, given by
\begin{equation}
\label{eq:dist2}
d=\ell_0\left(\mu^3-1\right)\, ,
\end{equation}
with $\mu $ defined in Eq.~\eqref{eq:parameters}. Two conditions are notable, namely (A) when LC1 bounces precisely $n$ times in the system during the ramp or (B) when it bounces $n$ times and then reaches $x=L/2$ at $ t=\tau $. These conditions are met, respectively, for $\tau=\tau^{\mathrm{(A)}}_{n}$ and $\tau=\tau^{\mathrm{(B)}}_{n}$, given by
\begin{eqnarray}
\tau^{\mathrm{(A)}}_{n}&=&\frac{3(\mu+1)}{2(\mu^2+\mu+1)}\frac{L}{v_i}n\, ,\\
\tau^{\mathrm{(B)}}_{n}&=&\frac{\tau_{1}^{(A)}}{2}+\tau_{n}^{(A)}\, .	
\end{eqnarray}
Note that, since $\tau^{\mathrm{(A)}}_1=(\mu^3-1)\tau_{\mathrm{ad}}/\pi$, unless for very weak quenches the adiabatic condition in Eq.~\eqref{eq:adcondition}  yields that LC1 bounces at least once in the system.
\begin{figure}[htbp]
	\begin{center}
		\includegraphics[width=\columnwidth]{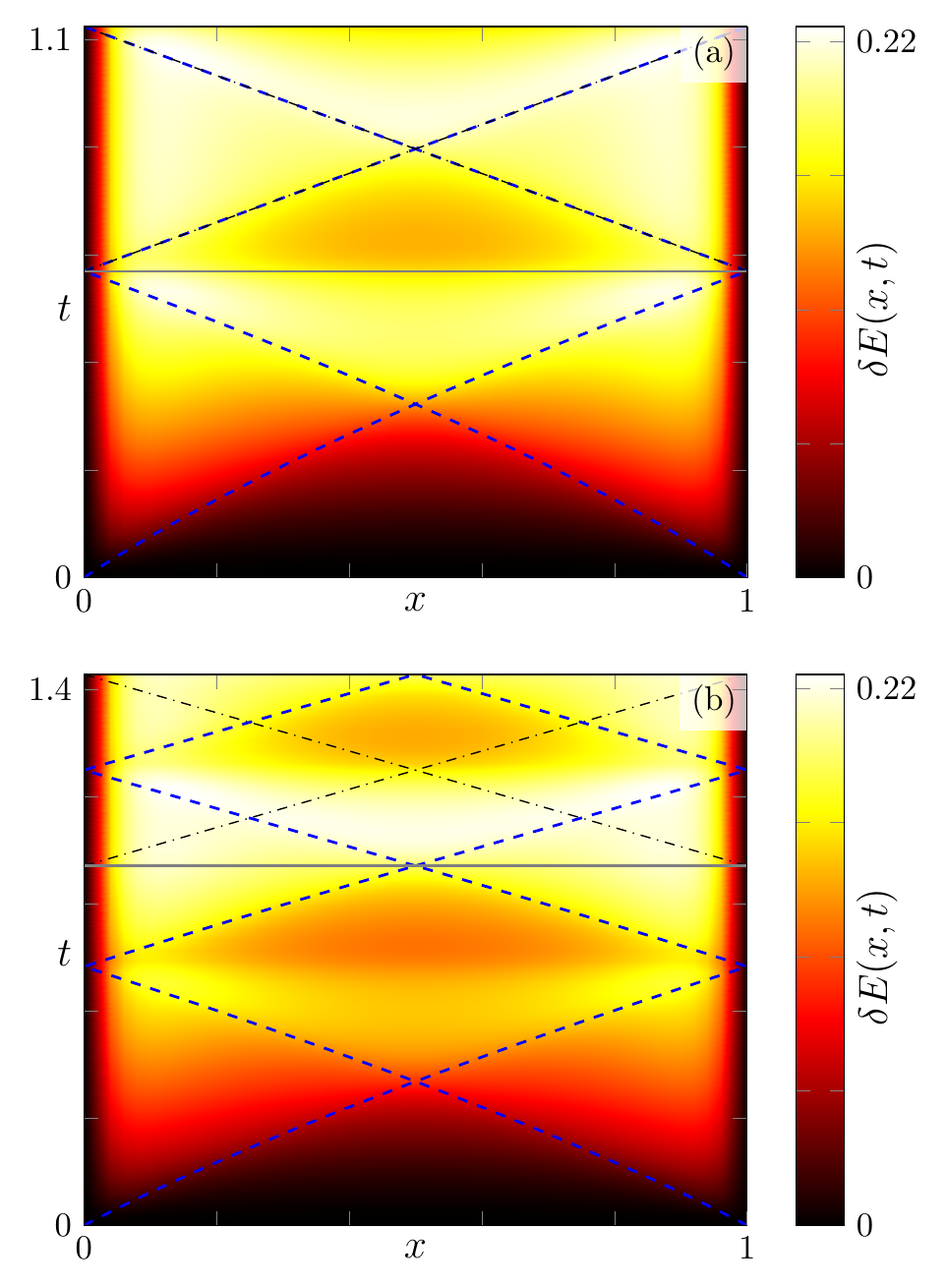}
		\caption{(Color online) Density plot of $ \delta E(x,t)=E(x,t)-E_{\mathrm{nq},i}(x) $ as a function of $ x $ (units $ L $) and $ t $ (units $ L/v_F $) for a quench from $ K_i=0.9 $ to $ K_f=0.5 $. The dashed blue lines represent LC1, while the dash-dotted black ones highlight LC2. The solid gray line denotes the end of the ramp. Ramp time: $ \tau=\tau^{\mathrm{(A)}}_1 $ (Panel (a)) and $ \tau=\tau^{\mathrm{(B)}}_1 $ (Panel (b)). Here, $ \alpha/L=0.05 $.}
		\label{fig:Fig2}
	\end{center}
\end{figure}

\noindent  The dynamics described above is illustrated in Fig.~\ref{fig:Fig2}. When $\tau=\tau_{n}^{\mathrm{(A)}}$, LC1 and LC2 propagate together after the quench (see Panel (a)), while when $\tau=\tau_{n}^{\mathrm{(B)}}$ LC1 and LC2 begin their post-quench evolution being maximally distant (see Panel (b)).\\
\noindent In the adiabatic regime, using Eqs.~(\ref{eq:DIIad}, \ref{eq:envelope}) and expanding to lowest order in the small parameter $L/\ell_0$ one can obtain an approximate analytic expression for the envelope function,
\begin{align}
	E(x,t)&\approx\left[\mathcal{K}(x)\right]^{\bar{K}(t)}\left[1-\frac{(1-\mu^{-1})L\bar{K}(t)}{24\pi\ell_0\mu^2}C(x,d)\right]\nonumber\\
	&\times\left[1-\frac{L\bar{K}(t)}{12\pi \ell_0}C(x,\ell(t))\right],
\end{align}
with $ \mathcal{K}(x) $ defined in Eq.~\eqref{eq:K}, $\bar{K}(t)=K_{i}/\sqrt{1+\eta t/\tau}$ (with $\bar{K}(\tau)=K_{f}$) and $C(x,y)$ defined in Eq.~\eqref{app:eq:C}. The term containing $C(x,\ell(t))$ describes the motion of LC1 through the system. Indeed, from the structure of Eq.~\eqref{app:eq:C} emerges that two perturbations propagate to the left and to the right with a law of motion precisely dictated by $\ell(t)$, as anticipated. A most notable feature here is the presence of an effective {\em instantaneous} value of the interaction parameter $\bar{K}(t)$, which interpolates between $K_i$ at $t=0$ and $K_{f}$ at $t=\tau$.\\

\begin{figure}[htbp]
	\begin{center}
		\includegraphics[width=\columnwidth]{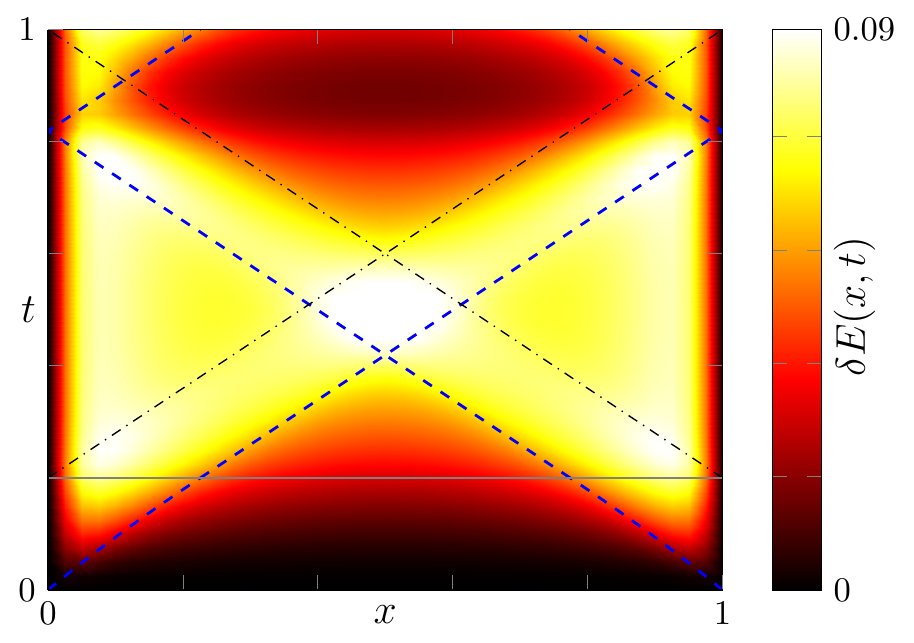}
		\caption{(Color online) Same as in Fig.~\ref{fig:Fig1} for a quench from $ K_i=1 $ to $ K_f=0.8 $ and ramp time $ \tau=0.2 $. The interplay between LC1 and LC2 is clearly visible. }
		\label{fig:Fig5}
	\end{center}
\end{figure}

\noindent (2) {\em{Post-quench}}. Employing Eqs.~(\ref{eq:DIIIad}, \ref{eq:envelope}, \ref{app:eq:EnvGGE}) one can obtain an approximated analytical expression for the envelope function in the post-quench region
\begin{align}
	\label{eq:LC12}
	E(x,t)&\approx E_{\mathrm{ad}}^{\mathrm{(GGE)}}(x)\left[1-\frac{LK_{f}}{12\pi\ell_0}C(x,v_{f}(t-\tau)+d)\right.\nonumber \\
	&\left.+\frac{LK_{f}^4}{12\pi\ell_0 K_{i}^3}C(x,v_{f}(t-\tau))\right]\, ,
\end{align}
with $E_{\mathrm{ad}}^{\mathrm{(GGE)}}(x)$ given in Eq.~(\ref{app:eq:EADGGE}). Here, all expressions have been expanded up to the first order in $L/\ell_0$. Two light cones now emerge, both propagating at constant speed $v_{f}$. One is a ``ghost" of LC1 and originates from where the latter was at $t=\tau$, while the second is LC2 and originates from the edges of the system when the quench ramp is finished. The amplitude of LC2 with respect to LC1 is modulated by the factor $K_{f}^3/K_{i}^3$. As a consequence, for a quench with $ K_f<K_i $, LC1 predominates and the best condition to observe both LC1 and LC2 for $t\geq\tau$ is that the quench is rather weak, i.e. $K_{f}\sim K_{i}$. The presence of two light cones is clearly visible in Fig.~\ref{fig:Fig5}, which shows a weak quench from $ K_i=1 $ to $ K_f=0.8 $ and ramp time $ \tau=0.2 $.\\

\noindent  From Eq.~\eqref{eq:LC12} one can notice that the time-dependent factors modulate $E(x,t)$ around the static envelope function $E_{\mathrm{ad}}^{\mathrm{(GGE)}}(x)$. This is the lowest order expansion in $L/\ell_0$ of the envelope function of the GGE distribution evaluated in Eq.~\eqref{app:eq:EnvGGE}. Therefore, $E(x,t)$ (and thus the density profile) fluctuates around the envelope obtained within the GGE approach, with an amplitude governed by the ratio $L/\ell_0$. The amplitude of such fluctuations then decreases as the quench is more adiabatic. Although in a finite system no steady distribution can be achieved after a quench, it however seems that the more LC1 can travel along the system, the larger a sort of effective thermalization occurs.\\
\begin{figure}[htbp]
	\begin{center}
		\includegraphics[width=\columnwidth]{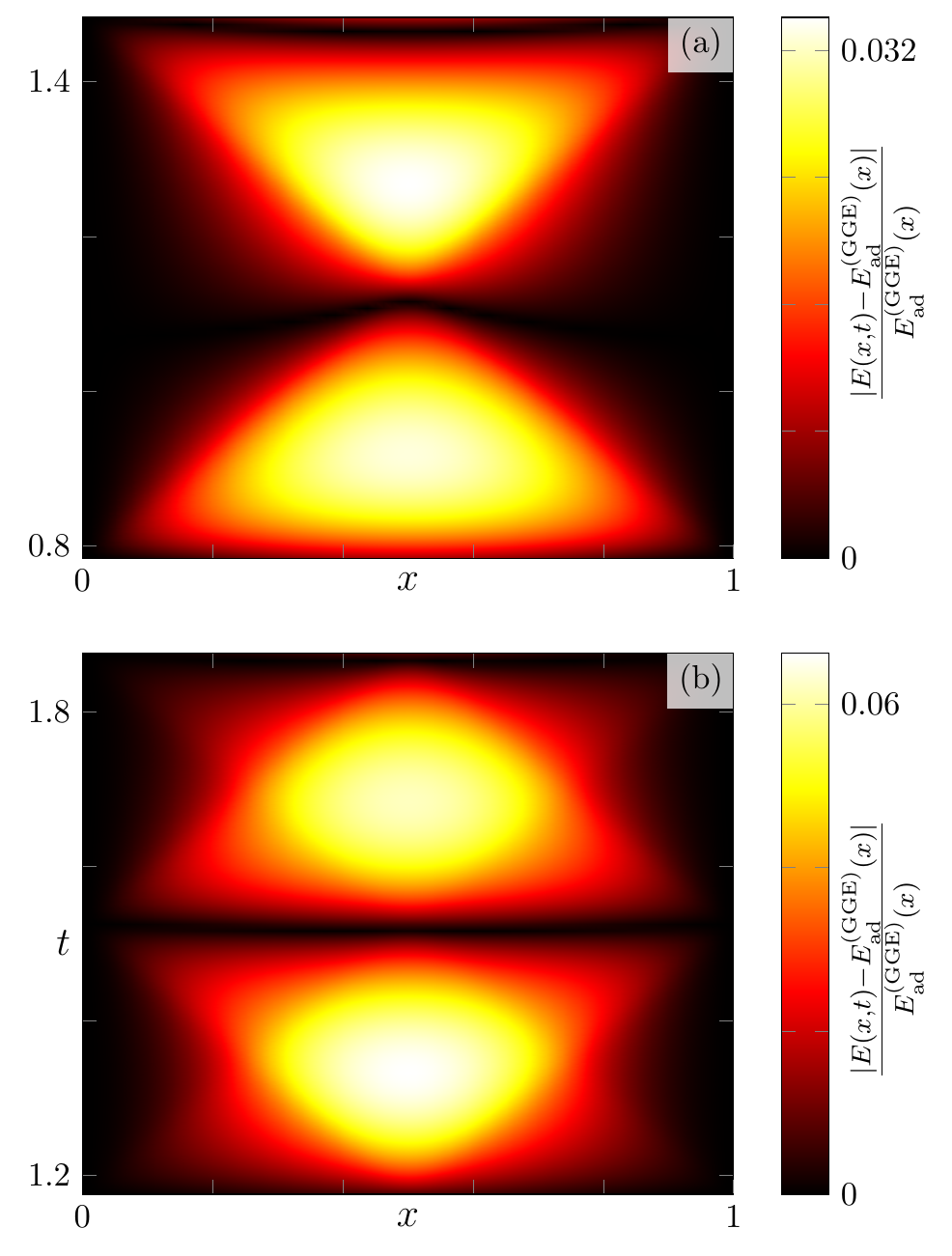}
		\caption{(Color online) Density plot of the relative difference $|E(x,t)-E_{\mathrm{ad}}^{\mathrm{(GGE)}}(x)|/E_{\mathrm{ad}}^{\mathrm{(GGE)}}(x)$ as a function of $ x $ (units $ L $) and $ t $ (units $ L/v_F $) for a quench from $ K_i=0.9 $ to $ K_f=0.7 $. Ramp time: $ \tau=\tau^{\mathrm{(A)}}_1 $ (panel (a)) and $ \tau=\tau^{\mathrm{(B)}}_1 $ (Panel (b)). Here, $ \alpha/L=0.05 $.}
		\label{fig:Fig4}
	\end{center}
\end{figure}

\noindent Figure~\ref{fig:Fig4} shows the relative difference $|E(x,t)-E_{\mathrm{ad}}^{\mathrm{(GGE)}}(x)|/E_{\mathrm{ad}}^{\mathrm{(GGE)}}(x)$ as a function of $x$ and $t$ over a period $\mathcal{T} $, for different values of $\tau$. As a general feature, comparing the situation for $\tau=\tau_{n}^{\mathrm{(A)}}$ and $\tau=\tau_{n}^{\mathrm{(B)}}$ shows that in the case of an integer number of bounces of LC1 in the system the envelope $E(x,t)$ is overall closer to the GGE limit with respect to the case when half more bounce occurs, when the distance is overall largest. This fact can be explained in terms of a destructive interference between LC1 and LC2 when they overlap. Indeed, since they enter in Eq.~\eqref{eq:LC12} with an opposite sign, their superposition lead to an overall suppression of the perturbation over the GGE envelope. 

\noindent It can also be seen that around the system edges $E(x,t)$ approaches almost perfectly $E_{\mathrm{ad}}^{\mathrm{(GGE)}}(x)$, as predicted for the general case below Eq.~\eqref{app:eq:EnvGGE}. 

\noindent When $\ell_0\lesssim L$, the approximation of Eq.~(\ref{eq:LC12}) ceases to be valid. In this regime, it is difficult to obtain analytic expressions and one has to resort to the numerical evaluation of Eq.~\eqref{eq:envelope}. Numerical analysis shows that as the adiabatic regime is left the difference between $E(x,t)$ and $E^{\mathrm{(GGE)}}(x)$ increases monotonously until the sudden quench regime is reached.\\

\subsection{Comparison with the GGE}
As seen above, in general $E(x,t)$ fluctuates around a static envelope function obtained assuming a fictitious steady distribution for the system in the GGE. On the other hand, a naive expectation would be that once the quench is performed, the density would tend to oscillate around the one corresponding to the final Hamiltonian, i.e. that the envelope function would be $E(x,t)\approx E_{\mathrm{nq},f}(x) $. In general, this is not the case. However, as can be seen in Eq.~(\ref{app:eq:EADGGE}), in the adiabatic limit $E_{\mathrm{ad}}^{\mathrm{(GGE)}}(x)\rightarrow E_{\mathrm{nq},f}(x)$. One can thus conclude that an adiabatic quench with a linear ramp may drive the system towards a state that weakly fluctuates around what essentially is the ground state distribution corresponding to the final Hamiltonian. On the other hand, the situation is completely different in the limit of a sudden quench: here $E_{\mathrm{sq}}^{\mathrm{(GGE)}}(x)$ has the same power-law form of $E_{\mathrm{nq},f}(x)$ with however a very different exponent -- see Eq.\eqref{app:eq:ESQGGE}.
\begin{figure}[htbp]
	\begin{center}
		\includegraphics[width=\columnwidth]{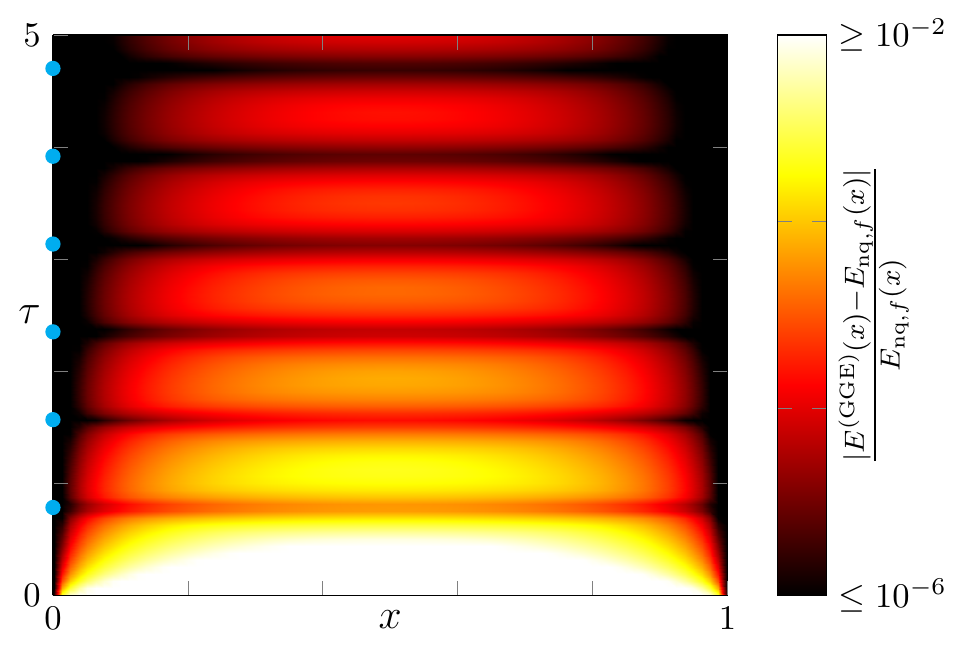}
		\caption{(Color online) Density plot of the relative difference $|E^{\mathrm{(GGE)}}(x)-E_{\mathrm{nq},f}(x)|/E_{\mathrm{nq},f}(x)$ as a function of $ x $ (units $ L $) and $ \tau $ (units $ L/v_F $) for a quench from $ K_i=0.9 $ to $ K_i=0.7 $.  The blue dots on the $ y- $axis represents the sequence $ \tau^{\mathrm{(A)}}_{n} $, with $ n\in\{1,...,5\} $. Here, $ \alpha/L=0.05 $. }
		\label{fig:Fig8}
	\end{center}
\end{figure}

\noindent Figure~\ref{fig:Fig8} shows the relative difference $|E^{\mathrm{(GGE)}}(x)-E_{\mathrm{nq},f}(x)|/E_{\mathrm{nq},f}(x)$ as a function of $x$ and $\tau$. The overall picture confirms the trend sketched in the discussion above. It is suggestive to notice that as $\tau$ is increased, the relative difference does not decrease monotonically but exhibits an oscillatory behavior, with local minima corresponding to $\tau=\tau_{n}^{\mathrm{(A)}}$ and local maxima located at $\tau=\tau_{n}^{\mathrm{(B)}}$, further confirming the importance of the light cones dynamics in defining the post-quench behavior.

\section{Conclusions}\label{sec:conclusion} 
We have studied the time dynamics of the particle density of a spinless open-boundary fermionic LL during and after a finite duration quench of the inter-particle interaction. Due to its finite size, the system never reaches a steady state but continues to oscillate around a certain configuration. In the adiabatic quench limit, the dynamics of the particle density is strongly affected by the presence of {\em two} light-cone perturbations, one originating as soon as the quench begins and the other arising when the quench stops. These perturbations originate from the edges and propagate ballistically through the system with the instantaneous velocity of the LL bosonic excitations, bouncing elastically whenever they reach an edge. For peculiar values of the length of the quench ramp, the two light cones can interfere destructively and the particle density maximally approaches the behavior predicted by a GGE calculation throughout the entire system. Furthermore, for adiabatic quenches the density evaluated in the GGE picture tends to the electron density of the ground state of the final Hamiltonian. In the sudden quench limit, on the other hand, the two light cones are indistinguishable, leading to maximal discrepancy with respect to the GGE limit in the central region of the system. Regardless the length of the quench ramp, the exact density profile at the edges always closely matches the density calculated within the GGE. 

\begin{acknowledgements}
We acknowledge financial support by the DFG (SPP1666 and SFB1170 ``ToCoTronics'') (N. T. Z.) and of project MIUR-FIRB-2012-HybridNanoDev (Grant No. RBFR1236VV) (S. P., F. M. G., F. C., M. S.).
\end{acknowledgements}

\appendix
\section{Generalized Gibbs ensemble}\label{subsec:GGE}
In this Appendix we evaluate the average fermionic density in the GGE framework~\cite{gge,expgge,qqreview}. Following this approach, one can introduce a GGE with associated density matrix
\begin{equation}
\label{eq:rhogge}
\rho_{\mathrm{GGE}}=\frac{1}{Z_\mathrm{GGE}}e^{-\sum_q\lambda_qI_q},
\end{equation}
with $ Z_{\mathrm{GGE}}=\text{Tr}\{\exp(-\sum_q\lambda_qI_q)\} $ and $ \{I_q\} $ a certain set of conserved (and independent) integral of motions (but not necessarily all the possible ones). In our case of a LL subject to a quench with finite time duration $ \tau $, the values of the Lagrange multipliers $ \lambda_q $ can be determined by imposing that, after the transient, the averages of the integrals of motion $ \{I_q\} $ evaluated with the GGE are conserved and equal to their value at $ t=\tau $,
\begin{equation}
\langle I_q(\tau) \rangle_i=\langle I_q \rangle_{\mathrm{GGE}}=\frac{1}{e^{\lambda_q}-1},\qquad\forall q. 
\end{equation}
Following previous works on the sudden quench case~\cite{cazalilla,Iucci:2009}, the most natural choice for the integrals of motion is $ I_q=\alpha^{\dagger}_q\alpha_q $, with $ \alpha_q $ and $ \alpha^{\dagger}_q $ the bosonic operators that diagonalize the {\em final} Hamiltonian $ H_f=H(t\geq\tau) $. These new operators are connected to initial ones by the Bogoliubov transformation $ \alpha_q=\chi_+\beta_q+\chi_-\beta^{\dagger}_q, $ with $ \chi_\pm=(\mu^{-1/2}\pm\mu^{1/2})/2 $. Thus, from Eq.~\eqref{eq:betaevolution} one obtains 
\begin{equation}
\label{eq:Iq}
\langle I_q(\tau) \rangle_i\!=\mathcal{F}(\Delta_q,\mu)-\frac{1}{2},
\end{equation}
with
\begin{equation}
\mathcal{F}(\Delta_q,\mu)=\frac{\pi^2\Delta_q^2\mu^3}{12}\left[|\mathcal{C}(\Delta_q,\mu)|^2+|\mathcal{S}(\Delta_q,\mu)|^2\right].
\end{equation}
In the framework of the GGE we can now evaluate the average density assuming a LL with Luttinger parameter $ K_f $ prepared in a state described by the density matrix of Eq.~\eqref{eq:rhogge}. Following the same procedure outlined in Sec.~\ref{subsec:bosonization}, with the only substitution $ \beta_q\rightarrow\alpha_q $ in Eq.~\eqref{eq:phi}, one obtains for the average fermionic density in the GGE
\begin{equation}
\langle\rho(x)\rangle_{\mathrm{GGE}}\!=\!\frac{N}{L}\left\{\! 1\!-\!E^{\mathrm{(GGE)}}(x)\cos\left[2k_Fx-2f(x)\right] \!\right\},
\end{equation}
with 
\begin{equation}
\label{app:eq:EnvGGE}
E^{\mathrm{(GGE)}}(x)\!=\exp\!\Big[\!-\!\frac{4\pi K_f}{L}\sum_{q>0}\!\frac{e^{-\pi\alpha/L}}{q}\sin^2(qx)\mathcal{F}(\Delta_q,\mu)\Big].
\end{equation}
Since our system possesses a finite length and thus a discrete energy spectrum, an asymptotic steady state for $ t>\tau $ will never be reached and the average of a generic observable, such as the density, will continue to oscillate around a certain value with a recurrence time $ \mathcal{T} $~\cite{Bocchieri:1957}. Thus, we don't expect that the GGE approach can capture all the details of the density dynamics. However, in the post-quench region for $ x\ll v_ft $ and $ L-x\ll v_ft $, i.e. around the system edges, one can neglect the spatial parts in the dynamical term of Eq.\eqref{eq:envelope}, obtaining that $ E(x,t)\rightarrow E^{\mathrm{GGE}}(x) $.

{\em Sudden quench.} In the sudden quench limit, using Eq.~\eqref{eq:DIIISQ}, we can expand Eq.~\eqref{app:eq:EnvGGE} to the zeroth order in $ \Delta_q $ obtaining
\begin{equation}
\label{app:eq:ESQGGE}
E_{\mathrm{sq}}^{\mathrm{(GGE)}}(x)=\left[\mathcal{K}(x)\right]^{\frac{K_{f}^2+K_{i}^{2}}{2K_{i}}}.	
\end{equation}
Note that $ E_{\mathrm{sq}}^{\mathrm{(GGE)}}(x) $ has the same functional form of the envelope of a non-quenched LL -- see Eq.~\eqref{eq:Estd} -- with effective Luttinger parameter $ (K_{f}^2+K_{i}^{2})/2K_{i} $.

{\em Adiabatic quench.} On the other hand, using Eq.~\eqref{eq:DIIIad}, one obtains the following asymptotic expansion for $ t>\tau $ in the adiabatic quench limit, valid to the first order in $ \Delta_q^{-1} $,
\begin{equation}
\label{app:eq:EADGGE}
E_{\mathrm{ad}}^{\mathrm{(GGE)}}(x)=\left[\mathcal{K}(x)\right]^{K_f}\left[1-\frac{K_{f}L(1-\mu^{-1})}{24\pi\ell_0\mu^2}C(x,d)\right]\, .
\end{equation}
Here, $ d=2v_{i}\tau(\mu^3-1)/(3\eta) $ and 
\begin{equation}
\label{app:eq:C}
C(x,y)=2\mathcal{D}\left(y\right)-\mathcal{D}\left(y+x\right)-\mathcal{D}\left(y-x\right),
\end{equation}
with
\begin{equation}
\mathcal{D}\left(y\right)=\text{Im}\left[\text{Li}_2\left(e^{-\alpha \pi /L+2\pi i y/L}\right)\right],
\end{equation}
where $ \text{Li}_2(x) $ is the dilogarithm function. See Sec.~\ref{results} for further details. 



\begin{thebibliography}{99}
\bibitem{footnote:tauvalues} We have assumed a system of length $ L=100\ \mathrm{nm} $ with $ N=10 $ particles and chosen $ q_c=k_F $. For carbon nanotubes we use $ v_F=10^{6}\ \mathrm{m/s} $, while for cold fermionic gases we set $ v_F\sim1\ \mathrm{m/s} $.
\end{thebibliography}
\end{document}